\documentclass[preprint,12pt]{aastex}
\usepackage{natbib}

\def\rmmat#1{{\hbox{\rm #1}}}
\def\rmscr#1{\rmmat{\scriptsize #1}}
\newcommand{\beq}{\begin{equation}}
\newcommand{\eeq}{\end{equation}}
\newcommand{\bea}{\begin{eqnarray}}
\newcommand{\eea}{\end{eqnarray}}
\newcommand{\bean}{\begin{eqnarray*}}
\newcommand{\eean}{\end{eqnarray*}}
\newcommand{\ba}{\begin{array}}
\newcommand{\ea}{\end{array}}
\newcommand{\bml}{\begin{mathletters}}
\newcommand{\eml}{\end{mathletters}}
\newcommand{\ie}{{\it i.e.~}}

\def\figref#1{Figure~\ref{fig:#1}}

\def\d{{\rm d}}

\def\dd#1#2{\frac{\d #1}{\d #2}}

\begin{document}
\title{LMXBs may be important LIGO sources after all}

\author{Jeremy S. Heyl\altaffilmark{1}}
\affil{Harvard-Smithsonian Center for Astrophysics, Cambridge, 
MA 02138; jheyl@cfa.harvard.edu}
\altaffiltext{1}{Chandra Fellow}

\begin{abstract}

Andersson et al.\ and Bildsten proposed that the spin of accreting
neutron stars is limited by removal of angular momentum by
gravitational radiation which increases dramatically with the spin
frequency of the star.  Both Bildsten and Andersson et al. argued that
the $r-$modes of the neutron star for sufficiently quickly rotating
and hot neutron stars will grow due to the emission of gravitational
radiation, thereby accounting for a time varying quadrupole component
to the neutron star's mass distribution.  However, Levin later argued
that the equilibrium between spin-up due to accretion and spin-down
due to gravitational radiation is unstable, because the growth rate of
the $r-$modes and consequently the rate of gravitational wave emission
is an increasing function of the core temperature of the star.  The
system executes a limit cycle, spinning up for several million years
and spinning down in less than a year.  However, the duration of the
spin-down portion of the limit cycle depends sensitively on the
amplitude at which the nonlinear coupling between different $r-$modes
becomes important.  As the duration of the spin-down portion increases
the fraction of accreting neutron stars which may be emitting
gravitational radiation increases while the peak flux in gravitational
radiation decreases.  Depending on the distribution of quickly
rotating neutron stars in the Galaxy and beyond, the number of gravitational
emitters detectable with LIGO may be large.
\end{abstract}

\section{Introduction}

Accretion onto the surface of a neutron star can in principle spin up
the rotation of the neutron star until the spin frequency equals the
Kepler frequency of the inner edge of the disk.  In low-mass x-ray
binaries, the disk is thought to extend to stellar surface so the
maximal frequency that the neutron star can achieve exceeds 1 kHz.
However, the observed and inferred spin frequencies of neutron stars
in low-mass x-ray binaries (LMXBs) are clustered around 250--500 Hz
\citep[e.g][]{1998ApJ...501L..89B}; the millisecond X-ray pulsars
SAX~J1808.4 and XTE~J1751 have slightly higher frequencies of 402~Hz
and 435~Hz \citep{1998Natur.394..344W,2002IAUC.7867....1M}.
Millisecond radio pulsars have been discovered with frequencies up to
640~Hz \citep{1982Natur.300..615B}.  All of these limits are well
below the Keplerian limit on the spin frequency of a neutron star, so
an alternative explanation for the maximal observed spin frequency of
neutron stars is required.  \citet{1999ApJ...516..307A} and
\citet{1998ApJ...501L..89B} proposed that inertial modes (specifically
the $r-$modes) inside the neutron may grow while generating
gravitational waves (GW).  For sufficiently quickly rotating stars, GW
carry away the angular momentum as quickly as it is deposited on the
star by accretion.

\citet{1999ApJ...517..328L} found that this proposed equilibrium
between spin-up and spin-down is unstable.  A neutron star will
execute a limit cycle \citep[see][for additional
discussion]{2000ApJ...534L..75A}.  It spins up for several million
years and then quickly spins down emitting GW in less than a year.
Only a small fraction of neutron stars is spinning down at any time;
it is unlikely that any neutron star within the galaxy is currently
spinning down, so none would be detected by LIGO.  However, the
duration of the spin down depends sensitively on the assumed maximal
amplitude of the $r-$mode.  \citet{1999ApJ...517..328L} assume that
the $r-$modes saturate when their amplitude is of order unity.
\citet{2001astro.ph.10487S}, \citet{2001ApJ...549.1011W}
and \citet{Arra02} found that the
saturation amplitude may be two to three orders of magnitude smaller;
this may increase the duration of spin down to be greater than several
thousand years.  This dramatically increases the number of neutron
stars whose GW could be detected both by increasing the typical GW
flux relative to the estimates of \citet{1999ApJ...516..307A} and
\citet{1998ApJ...501L..89B} and the number of currently emitting
sources relative to \citet{1999ApJ...517..328L}.

This Letter explore the implications of saturation of $r-$modes in
rapidly rotating neutron stars whose spins are accelerated by
accretion.  \S\ref{sec:calc} will describe a series of straightforward
calculations similar to those of \citet{1999ApJ...517..328L} but with
various values of the saturation amplitude (\S\ref{sec:spincalc}).
Both \citet{2000ApJ...534L..75A} and \citet{2001IJMPhD.10..381A} have
estimated the number of observable sources if the duty cycle corresponds
to an $r-$mode saturation amplitude of unity.  The observed number 
of millisecond pulsars, the presumed descendents of LMXBS 
\citep[e.g.][]{1991PhR...203....1B},
yields an estimate of the number of potential
sources and their amplitudes (\S\S\ref{sec:distcalc2}-\ref{sec:distcalc3}) 
as a function of the duty cycle.  
\S\ref{sec:discuss} will outline some consequences of
these results and the observational outlook.

\section{Calculations}
\label{sec:calc}

\subsection{Thermal and spin evolution}
\label{sec:spincalc}

The calculations presented here are essentially a recapitulation of
those of \citet{1999ApJ...517..328L}; however, a brief summary of the
equations governing the system is useful \citep{1998PhRvD..58h4020O}.
The angular velocity of the star $\Omega$ is normalized by the
dynamical frequency of the star, ${\tilde \Omega}=\Omega/(\pi G {\bar
\rho})^{1/2}$ where ${\bar \rho}$ is the mean density of the star.  If
the dimensionless $r-$mode magnitude is less than the amplitude at
which the mode saturates $\alpha_\rmscr{max}$, the spin and $r-$mode
amplitude are determined by
\bea
\dd{\tilde \Omega}{t} &=& -\frac{2 \alpha^2 Q}{1+\alpha^2 Q}
\frac{\tilde \Omega}{\tau_\rmscr{v}} + \sqrt{\frac{4}{3}} \frac{1}{\tilde I}
\frac{\dot M}{M} p 
\label{eq:dodt} \\
\dd{\alpha}{t} &=& - \left ( \frac{1}{\tau_\rmscr{grav}} +
\frac{1}{\tau_\rmscr{v}} \frac{1-\alpha^2 Q}{1+\alpha^2 Q} \right ) \alpha
\eea
where $Q$ and ${\tilde I}$ are determined by the density profile of
the star, a $n=1$ polytrope \citep{1998PhRvL..80.4843L}.  
Following \citet{1999ApJ...517..328L}, $Q=0.094$ and
${\tilde I}=0.261$.  $p$ depends on the angular velocity at the inner
edge of the accretion disk relative to that of the star.  The time for
the star to spin up is inversely proportional to $p$, so it can be
subsumed into the uncertain value of the accretion rate ${\dot M}$,
\ie $p=1$.  ${\dot M}$ is assumed to be 
$10^{-8} \rmmat{M}_\odot \rmmat{yr}^{-1}$ and 
$M$ is the mass of the star.  The timescale for the growth
and attenuation of the $r-$mode for a fluid neutron-star consisting of 
neutrons, protons and electrons are 
$\tau_\rmscr{grav} = -3.26 {\tilde \Omega}^{-6} \rmmat{s}$
and $\tau_\rmscr{v}^{-1} = (1.03 \times 10^5 \rmmat{s})^{-1} T_8^{-2} +
(6.99 \times 10^{14} \rmmat{s})^{-1} T_8^6$
where $T_8=T/(10^8 \rmmat{K})$ \citep{1998PhRvD..58h4020O}.  

Both the growth and decay rates depend on the properties of the
neutron star.  The presence of a crust
\citep{2000PhRvD..62h4030L,2001ApJ...549.1011W,2001MNRAS.324..917L}, magnetic 
field \citep{2000ApJ...531L.139R},
hyperonic or superfluid core \citep{2000PhRvD..61j4003L,2002PhRvD..65f3006L} 
can dramatically affect the dynamics of $r-$modes in
accreting neutron stars, shrinking the
range of spin frequencies and temperatures where an $r-$mode will
grow. On the other hand, a strange-quark-matter core may decrease the
viscosity, shifting the instability region to cooler temperatures
\citep{2001astro.ph.11582A}.  Consequently, the observation of GW from
an accreting neutron star would provide a unique probe of the physics
of its interior.

If the mode is saturated ($\alpha \geq \alpha_\rmscr{max}$), $\d
\alpha/\d t = 0$ and
\beq
\dd{\tilde \Omega}{t} = \frac{2 {\tilde \Omega}}{\tau_\rmscr{grav}}
\frac{\alpha_\rmscr{max}^2 Q}{1-\alpha_\rmscr{max}^2 Q}
+ \sqrt{\frac{4}{3}} \frac{1}{\tilde I}
\frac{\dot M}{M} p .
\eeq
\citet{Arra02} provide a simple expression for $\alpha_\rmscr{max}$ in
terms of the total energy in the $r-$mode when nonlinear
couplings become important, $\alpha^2_\rmscr{max}=A / (\Omega
|\tau_\rmscr{grav}| {\tilde J})$.  The dimensionless constant $A$
\citep[$\alpha_e$ in][]{Arra02} reflects an inaccuracy in the matching
conditions between the coupled modes and may range from $\sim 10^{-3}$
to $\sim 1.$

A relation determining the thermal evolution of the system closes the
system of equations.  Neutrino emission by the modified URCA process
\citep{Shap83} is taken to dominate the cooling while the core is
heated through accretion and dissipation.  The equilibrium temperature
of the core under accretion alone is taken to be ${\bar T}=10^8$~K.
These assumptions yield the following equation
\beq
\dd{T}{t} = \frac{1}{C_v} \left [ \frac{\alpha^2 \Omega^2 M R^2
{\tilde J}}{\tau_\rmscr{v}} - 7 \times 10^{31} \left ( T_8^8 - {\bar
T}_8^8 \right ) \rmmat{ergs s}^{-1} \right ]
\eeq
where $C_v$ is the heat capacity of the star taken to be $1.4 \times
10^8 T_8~\rmmat{ergs K}^{-1}$ \citep{Shap83} and 
${\tilde J}=1.635 \times 10^{-2}$ for the density profile considered
\citep{1998PhRvL..80.4843L}. The mass of the star $M=1.4 \rmmat{M}_\odot$ 
and its radius $R=12.53$~km.

\begin{figure*}
\plotone{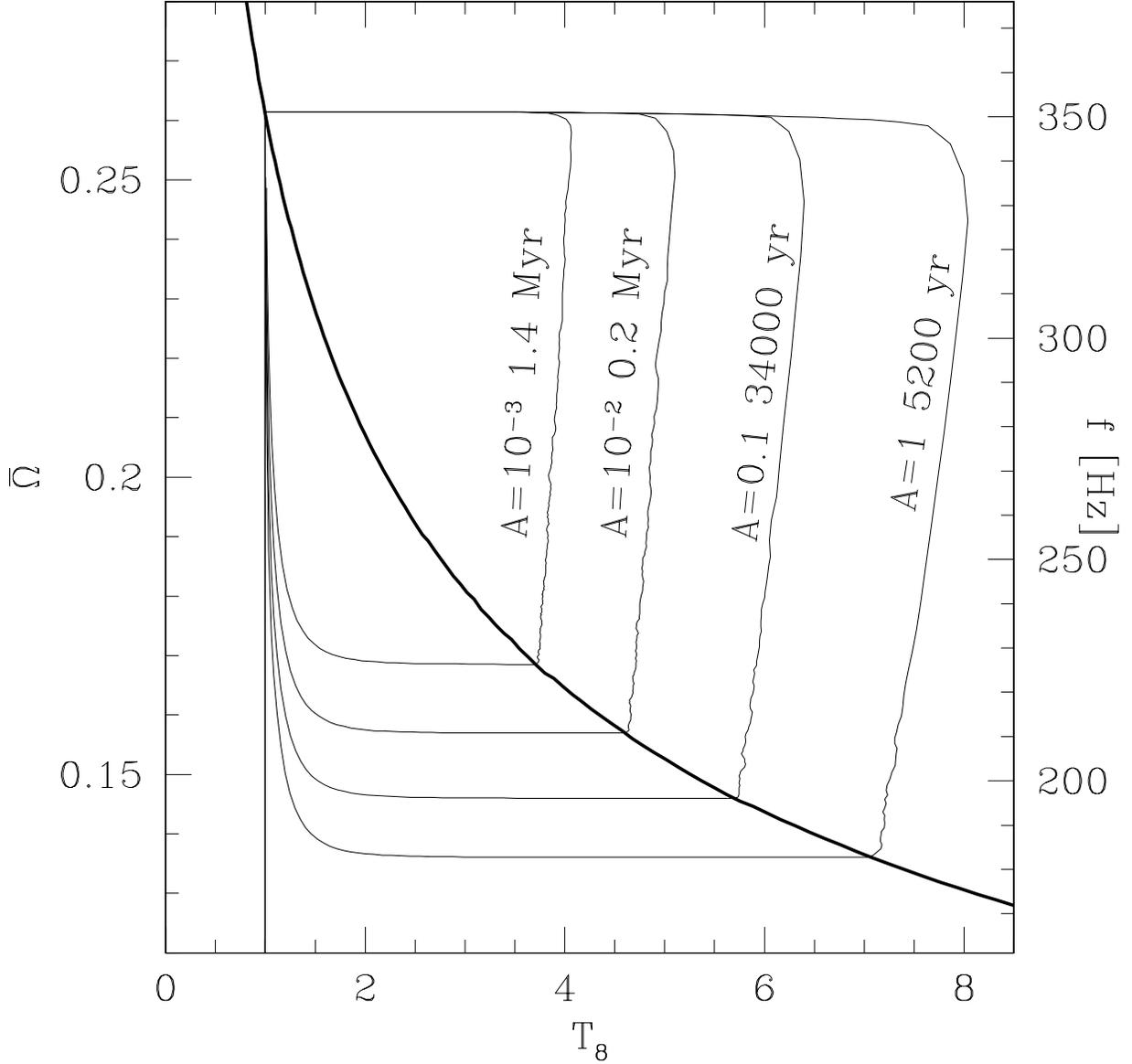}
\caption{ 
The limit cycle in core temperature and spin frequency 
for an accreting neutron
star with several values of the dimensionless matching constant,
$A$.  The star follows a clockwise trajectory in
phase space; from the outermost loop inward the values of
$A$ are 1, 0.1, 0.01 and $10^{-3}$.  The curves are
labelled with the duration of the GW emission during each circuit of
the limit cycle. The bold curve marks the boundary of the instability
region.  Above this curve, the amplification of the $r-$modes due to
gravitational wave emission dominates over their attenuation by
viscosity. }
\label{fig:limitcycle}
\end{figure*}

The calculation begins with $T_8=1$, ${\tilde \Omega}=0.1$ and
$\alpha=0$, whenever the system becomes unstable to growing $r-$modes,
$\alpha$ is set to $10^{-8}$.  \figref{limitcycle} depicts the limit
cycle for several values of $A$ which determines the amplitude at
which the mode saturates (for each value of $A$ the peak amplitude of
the mode is typically $2\times 10^{-3} A^{1/2}$).  Three important trends are
apparent.  The duration of the GW emission increases dramatically as
the saturation amplitude decreases.  The time for the system to
complete the circuit depends only weakly on $A$ (3.7--4.8 Myr), so the
duty cycle for GW emission also increases.  The range of spin
frequencies over the cycle decreases with $A$
\citep{2000ApJ...534L..75A}.  For this model the range in spin
frequencies over the population of LMXBs also depends on the assumed
equilibrium temperature of the core ${\bar T}$.  Stars which accrete
less on average will have smaller typical core temperatures, and
therefore experience a wider range of spin frequencies over the limit
cycle.  Finally, if the $r-$mode saturates at a smaller amplitude, the
temperature of the core during the epoch of GW emission decreases.
However, for $A>0.1$, the temperature of the core exceeds $5
\times 10^8$~K during spin down.

The relation is well fit by a power law with $\tau_\rmscr{on} \approx
5000 A^{-0.8} \rmmat{yr}$, in agreement with the discussion after
Eq. 91 of \citet{Arra02}.  The fraction of the time that a source is
emitting GW increases from $\sim 10^{-3}$ to $\sim 10^{-1}$ as $A$
decreases from unity to $10^{-3}$.  To probe the possible consequences
of a crust, superfluidity or other sources of additional shear viscosity, the
strength of the shear viscosity is increased by a factor of
sixty-four.  Here the critical spin frequency where the $r-$mode begins
to grow is doubled, the spin-down timescale decreases by a factor of
375, and the peak $r-$mode amplitude increases by a factor of 5.66.
Decreasing the equilibrium temperature of the core by a factor of two
(without changing the shear viscosity) has a less dramatic effect of
halving the spin-down timescale.  Depending on the properties of the
neutron star and how the $r-$mode amplitude saturates, the spin-down
timescale may vary from 14 to $10^{6}$ years; 34,000 years is the
result for a fluid star using $A=0.1$ \citep[the conservative estimate
of][]{Arra02}.

\subsection{Distribution of Sources - Galactic LMXBs}
\label{sec:distcalc2}

Obtaining a better estimate of the number of detectable GW sources
requires a estimate of the total number LMXBs in the Galaxy, their
accretion rates and distances.  Since the known LMXBs
number only about one hundred \citep{2001A&A...368.1021L}, one would expect
not to observe GW emission from any of them for duty cycles less than one 
percent.  Since LMXBs are thought to be the
progenitors of millisecond radio pulsars (MSPs), the demographics of
MSPs can provide an estimate of the total number of Galactic LMXBs
\citep{1982Natur.300..728A,1991PhR...203....1B}.  Equating the rate
of MSP formation to the rate of LMXB formation yields an estimate of
the number of LMXBs that could be emitting GW at any given time,
\beq
N_\rmscr{LMXB,on} = \frac{\tau_\rmscr{on}}{\tau_\rmscr{cycle}}
N_\rmscr{LMXB} = \frac{\tau_\rmscr{on}}{\tau_\rmscr{cycle}} (
r_\rmscr{MSP} t_\rmscr{LMXB} ) = \tau_\rmscr{on} \frac{{\tilde
\Omega}_\rmscr{MSP}}{\Delta {\tilde \Omega}} r_\rmscr{MSP} \sim
\frac{\tau_\rmscr{on}}{10^4 \rmmat{yr}}
\eeq
where $\tau_\rmscr{cycle}$ is the time for the neutron star to
traverse the limit cycle once.  This is essentially equal to the time
for the star to spin up by $\Delta {\tilde \Omega}$ from the bottom of
the limit cycle to the top.  $t_\rmscr{LMXB}$ is the time it takes the
neutron star to spin up as an accreting LMXB to the final spin rate
$\Omega_\rmscr{MSP}$, and $r_\rmscr{MSP}$ is the rate of formation of
MSPs in the Galaxy $\sim 3 \times 10^{-5}~\rmmat{yr}^{-1}$ 
\citep{1995Natur.376..393L,1998MNRAS.295..743L}.  The ratio of 
$\tau_\rmscr{cycle}$ to $t_\rmscr{LMXB}$ is inferred from Eq.~\ref{eq:dodt}.

Estimating the typical strain amplitude that would be observed from
one of these sources is also straightforward since the total angular
momentum radiated is proportional to $\Delta \Omega$.  This yields
\beq
h = \frac{\alpha}{8} \frac{R}{d} 
\left ( 10 \frac{{\tilde I} \Delta {\tilde \Omega} }{  < {\tilde
\Omega}  > } \frac{G M}{c^3 \tau_\rmscr{on}} \right )^{1/2} 
\approx 7 \times 10^{-25} \left ( \frac{M}{1.4 \rmmat{M}_\odot} 
 \frac{10^4 \rmmat{yr}}{\tau_\rmscr{on}} \right )^{1/2} 
\frac{R}{10 \rmmat{km}} \frac{10 \rmmat{kpc}}{d}
\eeq
where $\alpha$ is a constant which depends on geometry -- its mean
value is 2.9 \citep{Brad98}.  To obtain the final approximation,
$\Delta {\tilde \Omega}=0.12$ and $<{\tilde \Omega}>=0.2$ which is
appropriate for $A = 0.1$.  This is thirty times
larger than the strain for the brightest source, Sco~X-1, if the GW
emission is continuous \citep{1998ApJ...501L..89B} and should be
easily detected with the initial LIGO \citep{Brad98}. One could
possibly detect such sources even in M31, so for $\tau_\rmscr{on} \sim
10^4$~yr, one could expect to find several sources in the Local Group.

If $\tau_\rmscr{on}$ is greater than a few hundred years, the
radiation from the surface will reflect the heightened core
temperature \citep{1998PhR...292....1T}.  The X-ray radiation is powered
by the dissipation of the $r-$modes.  The typical core temperature as
the star is spinning down is $6 \times 10^8$~K, yielding a effective
temperature of $3 \times 10^6$~K and a soft-X-ray luminosity of
$10^{35}$~erg/s \citep{Heyl98numens}.  Assuming that the source is 
found at
the maximum possible distance for an enhanced LIGO with $h_\rmscr{min}
\sim 10^{-27}$ \citep{Brad98},
\beq
d_\rmscr{max} = 7~\rmmat{Mpc} 
\frac{10^{-27}}{h_\rmscr{min}} 
\left ( \frac{10^4 \rmmat{yr}}{\tau_\rmscr{on}} \right )^{1/2}
\left ( \frac{M}{1.4 \rmmat{M}_\odot}  \right )^{1/2}
 \frac{R}{10 \rmmat{km}}  .
\eeq
this yields an observed X-ray flux of
\beq
F = 2 \times 10^{-17}~\rmmat{erg cm}^{-2} \rmmat{s}^{-1}
\left ( \frac{h_\rmscr{min}}{10^{-27}} \right )^2
\frac{\tau_\rmscr{on}}{10^4 \rmmat{yr}}
\left ( \frac{M}{1.4 \rmmat{M}_\odot}  \right )^{-1}
\left ( \frac{R}{10 \rmmat{km}} \right )^{-2} 
\eeq
due to the dissipation of the $r-$mode energy alone.  The typical
Galactic source (at a distance of 10~kpc), discussed earlier would be
500,000 times brighter with a flux of about 3.2 mCrab.
\citet{2000ApJ...536..915B} discussed the X-ray emission produced if
the GW emission is steady and derive constraints on either the
accretion or the $r-$modes.  If the GW emission is transient, the
X-ray emission is significantly stronger while GW are being emitted but the
vast majority of LMXBs would not be in this stage, thereby avoiding these
constraints.

\subsection{Distribution of Sources - Extragalactic LMXBs}
\label{sec:distcalc3}

For much smaller values of $\tau_\rmscr{on}$, the expected number of active
sources in the Galaxy vanishes; however, any sources that are active
will be visible well beyond the Local Group.  The total number of
sources brighter than $h_\rmscr{min}$ is given by
\beq
N_\rmscr{LMXB,on} = \frac{4}{3} \pi \left (\frac{\alpha}{8}
\frac{R}{h_\rmscr{min}} \right )^3 \left ( 10 \frac{{\tilde I} \Delta
{\tilde \Omega} }{ < {\tilde \Omega} > } \frac{G M}{c^3} \right
)^{3/2} \tau_\rmscr{on}^{-1/2} \frac{{\tilde
\Omega}_\rmscr{MSP}}{\Delta {\tilde \Omega}} {\cal R}_\rmscr{MSP}
\eeq
where ${\cal R}_\rmscr{MSP}$ is the formation rate of MSPs per unit
volume averaged over the local region of the universe, ${\cal
R}_\rmscr{MSP} \approx r_\rmscr{MSP} {\cal L}/L_\rmscr{Milky Way}$ 
where ${\cal L}$ is 
the local luminosity density, about 
$2 \times 10^{-2} h_{100} L_\rmscr{Milky Way}$~Mpc$^{-3}$
\citep{Heyl97elf} where $h$ is the Hubble Constant in units of 
100~km/s/Mpc.  Using the values for $\alpha_\rmscr{max}=1$, the case 
examined by \citet{1999ApJ...517..328L}, yields,
\beq
N_\rmscr{LMXB,on} = 3 h_{100} 
\left ( \frac{10^{-27}}{h_\rmscr{min}} \right )^3
\left ( \frac{1 \rmmat{yr}}{\tau_\rmscr{on}} \right )^{-1/2}
\left ( \frac{M}{1.4 \rmmat{M}_\odot} \right )^{3/2}
\left ( \frac{R}{10 \rmmat{km}}  \right )^3 .
\eeq
The typical distance of the sources detected at this sensitivity with
$\tau_\rmscr{on} \sim 1$~yr would be nearly 1~Gpc.  Even if
$N_\rmscr{LMXB,on}$ is less than one, one would expect to discover one source
during a period of $\tau_\rmscr{on}/N_\rmscr{LMXB,on}$.

\section{Discussion}
\label{sec:discuss}

If the spin of accreting neutron stars is indeed limited by the
emission of gravitational radiation
\citep{1999ApJ...516..307A,1998ApJ...501L..89B}, low-mass X-ray
binaries may be an important source for LIGO.  Although neutron stars
may execute a duty cycle \citep{1999ApJ...517..328L} of spin-up and
spin-down which reduces the number of active sources at a given time,
the sources that are active are typically much brighter than in a
model where they emit constantly.  Unfortunately, the number of LMXBs
that have been discovered actively accreting is too small ($\sim 100$)
to determine which effect dominates for duty cycles less than ten
percent.

However, if low-mass X-ray binaries are assumed to be the exclusive
progenitors of millisecond pulsars \citep[see][ for
an alternative]{1984JApA....5..209V}, 
the demographics of the millisecond pulsars in the
Galaxy and beyond provides an estimate of the number of sources.
Specifically, if the duration of the epochs when the neutron star is
emitting gravitational radiation is greater than 10,000 years, several
objects in the Galaxy will be above the detection thresholds of LIGO.
For $\tau_\rmscr{on} \sim$~10,000 year, these sources would be
detectable throughout the Local Group.  Those objects in the Galaxy
could also be detected from their X-ray emission which would be
powered by gravitational radiation reaction; they would be GW-powered
neutron stars.

For $\tau_\rmscr{on}$ much less than 10,000~years and much greater
than one year, no sources are likely to be detectable even with an
enhanced LIGO detector.  However, if the duration of gravitational
wave emission per cycle is less than several years, several sources
could be detected by an enhanced LIGO.  In this case, the sources will
be located at a typical distance of 1 Gpc.

Depending on the nature of the $r-$mode instability in the cores of
quickly spinning neutron stars, LMXBs may provide gravitational-wave
beacons throughout the Galaxy and the Local Group or at cosmologically
significant distances.

\acknowledgements

I would like to thank Phil Arras for useful discussions, and
I was supported by the Chandra Postdoctoral Fellowship Award 
\#PF0-10015 issued by the Chandra X-ray Observatory Center, which is
operated by the Smithsonian Astrophysical Observatory for and on
behalf of NASA under contract NAS8-39073.

\bibliographystyle{apj}
\bibliography{ns,mine,wd2,rmode,lmxb}
\end{document}